\documentstyle[11pt,paspconf]{article}

\begin{document}

\title{Morphology and Stellar Populations in the Gas-Rich, Giant LSBGs
    }

\author{P. Knezek\altaffilmark{1}}
\affil{Center for Astrophysical Sciences, Bloomberg Center, 
    The Johns Hopkins University,
    Baltimore, MD 21218-2695}






\begin{abstract}
An unexpected characteristic of 
low surface brightness galaxies (LSBGs) is that a
significant number are massive and possess substantial amounts of atomic gas.
We present preliminary results of an ongoing program to obtain 
BVRIJHK imaging, along with some nuclear spectroscopy, 
of a well-defined sample of LSBGs which are gas-rich and of similar
size to giant, high surface brightness spiral galaxies (HSBGs).  
These LSBGs span the entire range of Hubble disk morphologies.
While their disks are bluer, on average, than comparable HSBGs, the
optical morphology of massive LSBGs indicates that many of these systems have 
undergone previous star formation episodes.  They typically have long
disk scale lengths, and range from 
M$_{\rm B} = -16$ to $-22$ (H$_{0} = 75$ km s$^{-1}$ Mpc$^{-1}$).  
About half of the LSBGs with bulges show evidence of nuclear activity, 
and $\sim$30\% 
have appear to be barred.  These massive, gas-rich LSBGs apparently have 
varied, and often complex, evolutionary histories.

\end{abstract}


\keywords{low surface brightness galaxies}


\section{Introduction}

One of the discoveries in the work on low surface 
brightness galaxies (LSBGs) was that a significant number are 
massive and possess
significant quantities of atomic gas.  Many disk galaxies with sizes 
approaching that of Malin 1 (M$_{\rm HI}$ $\sim$ 2x10$^{11}$ M$_{\odot}$; D $\sim$ 147 kpc, 
H$_{0}$ = 75 km s$^{-1}$ Mpc$^{-1}$; Impey and Bothun 1989) 
have been identified through the use of the Palomar 
Sky Surveys (Schombert et al.\ 1992) and UK Schmidt plates (Impey et al.\ 
1996), with follow-up studies in HI.
While many of these 
galaxies have large amounts of atomic gas and unusually blue disk colors, they 
have only
weak regions of H$\alpha$ emission, indicating 
little ongoing massive star formation, and their low surface brightnesses 
suggest extremely low stellar surface densities.  
The blue disk colors have ruled out the hypothesis that all LSBGs are faded 
galaxies, but
the difficulty in
disentangling the difference between low metallicity and young stellar 
populations based on broadband optical colors has hindered understanding their 
stellar evolutionary history.
It has 
been suggested 
(McGaugh 1992) that some LSBGs may, in fact, be undergoing their first episodes 
of star formation since there is little evidence of a difference in the 
distribution of light in $UBVRI$ images.  
The LSBGs in his sample are largely 
{\it dwarf} systems, however, 
and the situation is not necessarily comparable for more massive LSBGs.  
In fact, as can be seen below, 
the optical morphology of massive LSBGs in $B$ and $R$ is consistent 
with 
the idea that many of these systems have undergone previous episodes of star
formation.  

We are conducting an extensive survey of both broadband optical ($BVRI$) 
and near-infrared ($JHK$) imaging of a 
sample of LSBGs which are gas-rich 
(M$_{HI}\geq$\kern.03em 5x10$^{9}$\kern.03emM$_{\odot}$) and of similar size 
to giant spiral galaxies (D$_{25}\geq$\kern.03em25\kern.03emkpc).  
We have also obtained spectroscopy of the nuclear regions of a small sample
of HI-rich giant LSBGs which indicate that some LSBGs, particularly
those with bulges, {\it do} have low level active galactic nuclei (Knezek
\& Schombert 1999).

\section{The Sample}

The primary sample consists of galaxies identified from the {\it Uppsala 
General Catalog of Galaxies} (Nilson 1973) based on the first Palomar 
Observatory Sky Survey, and supplemented by galaxies from the second 
Palomar Sky Survey (Schombert \& Bothun 1988; Schombert et al.\ 
1992).  These galaxies are selected to be gas-rich
(M$_{HI}\geq$\kern.03em 5x10$^{9}$\kern.03emM$_{\odot}$) and of similar size
to giant spiral galaxies (D$_{25}\geq$\kern.03em25\kern.03emkpc), assuming 
H$_{\rm o}$ $=$ 75 km s$^{-1}$ Mpc$^{-1}$.  No morphological criteria is 
applied other than the galaxies must be identified as disks.  Atomic hydrogen 
data is from Schneider et al.\  (1990, 1992).  Only galaxies with 
galactic latitudes $> \pm$30\deg\ are included.  Galaxies are then 
separated into ``high'' and ``low'' surface brightness samples based on 
their mean blue surface brightnesses, which are initially 
estimated from published 
blue magnitues and sizes from either Nilson (1973) or Schombert et al.\ 
(1992), then refined using our own data.  
Those galaxies with $\mu_{\rm B} < 24.5$ mag arcsec$^{-2}$ are 
designated giant high surface brightness galaxies (HSBGs), and those with 
$\mu_{\rm B} > 24.5$ mag arcsec$^{-2}$ are designated giant low surface 
brightness galaxies (LSBGs).  This is a mean blue surface brightness 
measured within $\mu_{\rm B} = 26.0$ mag arcsec$^{-2}$ from our optical data.

\section{Observations}

The broadband optical and near-infrared (NIR) imaging presented here are 
part of a larger, ongoing, project to image the entire sample of 175 LSBGs
selected according to the criteria above, as well as a sample of 
corresponding HSBGs to use for comparison.  Imaging has primarily 
been accomplished through the use of Michigan-Dartmouth-MIT Observatory
(MDM; 1993-1995),
San Pedro Martir Observatory (1996-present), and Las Campanas Observatory
(1995-1997).  A few
observations of galaxies too large to be imaged at those facilities were
obtained using the Kitt Peak National Observatory 0.9m in 1993.  
$BVRI$ imaging has been obtained of $\sim$ half the sample of 175 LSBGs, and
$JHK$ imaging has been obtained for $\sim$ one-third of the sample.
These data have been corrected for galactic extinction using the interpolation 
program provided by Burstein and Heiles.  No intrinsic extinction correction 
is applied.
Spectroscopic observations of a sample of HI-rich galaxies were
obtained using the MDM 2.4m in September, 1992 and January, 1993 
(Knezek \& Schombert 1999).

\section{Discussion}

As can be seen in Figure 1, the most obvious trait of these gas-rich
LSBGs is that their optical morphology spans the {\it entire} Hubble
Sequence for disk galaxies.  
\begin{figure}
\vspace{6in}
\caption{$B$ images of UGC~12740, UGC~2712, UGC~9022, and UGC~10313.}  \label{fig1} 
\end{figure}
{\it Not all LSBGs are late-type disk galaxies.}  Furthermore, it is not
possible to determine whether a disk galaxy is a dwarf or a giant simply
by its optical morphology.  Many of these gas-rich, massive LSBGs were
originally classified as dwarfs by Nilson (1973).  Their optical morphologies
are often indistinguishable from the true ``dwarf spirals''  studied by
Pildis et al.\ (1997) and Matthews \& Gallagher (1997).  Yet the average
disk scale length for an LSBG from our sample is $\sim$7 kpc, versus less
than 1 kpc for the ``dwarf spirals''.  Also, based on our optical 
morphology, $\sim$30\% of the LSBGs are barred systems, a comparable number
to that found for HSBGs.

We find 
that the gas-rich disk 
galaxies have very blue disks on average, whether they are
high or low surface brightness systems. LSBGs with a prominent bulge typically
have redder {\it disks}, with $<B-R> \sim 1.0$.  The bulges themselves have
$<B-R> \sim 1.4-1.6$, comparable to the bulge colors of HSBGs.  
Many of the bluest systems, 
with $<B-R> \sim 0.75$, can be
fit by ``pure disks'', i.e. require a single exponential component to
characterize their radial light distributions.  These systems 
are too blue to be explained by old, metal 
poor stellar populations.  Comparing to models by Worthey (1994), 
we find that even if the metallicity of a galaxy is only 1\% solar, stars
with ages of 8${\rm x}$10$^{9}$ years produce a system that is too red.
Yet most LSBGs have only a few HII regions, and metallicities that are
closer to 30\% solar (McGaugh 1994).  
Furthermore, preliminary results based on the addition of the NIR data
suggest that the colors of these systems are inconsistent with a Salpeter
IMF and a constant star formation rate.  Apparently, despite their low
stellar surface densities, many LSBGs have had a complex star formation
history.

Studies of the molecular emission of these LSBGs (Knezek 1993) indicate
that only the redder LSBGs, and those with bulges, possess measureable 
molecular hydrogen.  Furthermore, it is only the redder, gas-rich
LSBGs with bulges that have evidence of nuclear activity (Knezek \&
Schombert 1999).  Of those which do exhibit nuclear activity, over half
have line ratios indicative of active galactic nuclei (AGNs) rather than
star formation.  Those with AGNs show no evidence of a broad line region.
Finally, we find that there is a cut-off in the 
relationship between disk central surface brightness and disk scale length, in
the sense that {\it no} galaxies with bright disk central surface brightnesses 
have long disk scale lengths.  These correlations may provide a 
clues to the formation and evolution of these LSBGs, and the underlying 
physics of star formation in disk galaxies in general.

\section{\bf Conclusions}
Results of our ongoing project indicate that:

\bigskip
\par\noindent
$\bullet$ LSBGs have optical morphologies which span the entire range of
Hubble disk types.  It is {\it not} possible to determine whether a disk 
galaxy is a dwarf, in luminosity and mass, simply by it's optical morphology.
Furthermore, some bulge-dominated LSBGs have likely been missed simply 
because sample selections generally use mean surface brightnesses or late-type
morpohologies as a criteria.

\vskip0.25truein
\par\noindent
$\bullet$ A significant fraction of the massive 
LSBGs ($\sim$30\%) appear to possess
optical bars.

\vskip0.25truein
\par\noindent
$\bullet$ Gas rich, massive disk galaxies have very blue disk colors on
average.  For those galaxies with bulges, the bulge colors are normal.  
Preliminary results based on the additon of optical and 
near-infrared data suggests that the colors are inconsistent with a
Salpeter IMF with a constant star formation rate.  
We have compared our
data to models of the evolution of single burst stellar populations
with varying ages and metallicities, assuming a standard IMF
(Worthey 1994).  If we assume
metallicities of 30\% solar and an age typical for globular clusters,
$12\times 10^{9}$ years, the resultant color is $B-R=1.43$.  If we
assume the most extreme case, where the metallicity is 1\% solar and
the age is only $8\times 10^{9}$ years, the modelled color is
$B-R=1.04$.  While this is consistent with our average disk color
for the sample as a whole, it is still too red for the pure disk
systems, which have colors of $B-R=0.749\pm 0.166$.

\vskip0.25truein
\par\noindent
$\bullet$ Only the redder, early-type LSBGs with bulges show evidence
for molecular gas and AGN activity.

\acknowledgments

I am grateful to S. Lawrence and I. Cruz-Gonzalez for
their continuing help with the observations of this ongoing project.

\end{document}